\documentclass{aastex}     
\usepackage{spr-astr-addons}  
\usepackage{url}
\usepackage{comment}
\usepackage{epstopdf}
\usepackage{gensymb}
\usepackage{xcolor}

\usepackage[maxfloats=40] {morefloats} 

\begin{document}
\title{ Transit modelling of selected {\em Kepler} systems}
\shorttitle{Exoplanet Transits and Uncertainties}
\shortauthors{<Huang Q.Y.  et al.>}

\author{Q.~Y.\ Huang\altaffilmark{1}}
\and
\author{T.\ Banks\altaffilmark{2}}
\email{tim.banks\@nielsen.com}
\and
\author{E.\ Budding\altaffilmark{3,4,5,6}}
\and
\author{C.\ Puskullu\altaffilmark{3}}
\and
\author{M.~D.\ Rhodes\altaffilmark{7}}

\altaffiltext{1}{Munich Re, 20 Collyer Quay {\#}13-01, Singapore 049319}
\altaffiltext{2}{Nielsen, Data Science, 200 W Jackson Blvd \#17, Chicago, IL 60606, USA. Email: tim.banks@nielsen.com, Tel: 1-847-284-4444}
\altaffiltext{3}{Canakkale Onsekiz Mart University, Physics Department, TR 17020, Turkey}
\altaffiltext{4}{Dept.\ Physics \& Astronomy, University of Canterbury, New Zealand}
\altaffiltext{5}{SCPS, Victoria University of Wellington, P.O. Box 600, Wellington, New Zealand}
\altaffiltext{6}{Carter Observatory, PO Box 893, Wellington 6140, New Zealand}
\altaffiltext{7}{BYU, Provo, Utah, USA}

\vspace{2mm} 
\begin{abstract}

This paper employs a simple model, considering just geometry and linear
{  or quadratic} limb darkening, to fit {\em Kepler} transit
data via a Markov Chain Monte Carlo (MCMC) methodology for Kepler-1b,
5b, 8b, 12b, 77b, 428b, 491b, 699b, 706b, and 730b.  Additional fits
were made of the systems using the more sophisticated modeller { \em
Winfitter}, which {  gives} results in general agreement with
the simpler model. Analysis of data with longer integration times showed
biasing of the derived parameters, {  as expected from the
literature,} leading to larger estimates for radii and reducing
estimates of the system inclination. 
    
\end{abstract}

\keywords{optimization; exoplanets; light curve analysis}


\section{Introduction}

Since the first discoveries of planets orbiting other stars over two
decades ago, many thousands have been discovered (see Pollacco {\em et
al}.\ (2006) and Rice (2014) for reviews). Transits, where the planet
passes in between its host star and an observer's line of sight leading
to a dimming which can be modelled, have been the main data source for
exoplanet detection to date. The {\em Kepler} mission is currently the
major provider of such data (see Borucki {\em et al.}, 2003 and 2011,
for further information on this mission). The Kepler Science Center has
managed the organization of these  data for scientific users, being
readily available from the NASA Exoplanet Archive (NEA:
{http://exoplanetarchive.ipac.caltech.edu,}  {Akeson~{\em et al.},
2013}).  

This paper makes use of data from the NEA and focuses on
recovering the transit parameters with their uncertainties for
exoplanets.  We proceeded with these steps:
\begin{itemize}
\item Build a {  simple planetary transit light curve} model using the Python
programming language.

\item Fit the light curve model on five {  ``known'' (or test)} exoplanet data-sets
(Kepler 1b, Kepler 5b, Kepler 8b, Kepler 12b and Kepler 77b), for which
{  multiple} published results from transit modelling exist. {  Compare
the parameter estimates with the literature, to gain reassurance that
our model produces results similar to those from other methods, ideally
within confidence ranges.  These fits would be based on the
Levenberg-Marquadt algorithm, and provide starting parameter estimates
for Hamiltonian Monte Carlo (HMC) optimizations. Then compare the
results with those published elsewhere or as listed on the NEA website.

\item Fit the light curve model on systems without  { 
multiple fits in the literature,} using Markov Chain Monte Carlo (MCMC) procedures
to obtain  {  independent} estimates and uncertainties of the
parameters for the systems. Kepler 428b, Kepler 491b, Kepler 699b,
Kepler 706b, and Kepler 730b {  were selected on the basis of
having deep transits and uncomplicated light curves (e.g., visual
inspection showed no obvious ellipticity, strong reflections, single
planet, etc.) Morton {\em et al.'s} (2016) probabilistic validation
method tests all conceivable astrophysical false positive scenarios,
producing estimates whether the cause of a transit candidate signal is
likely due to planet transiting the presumed target star.  Morton
estimates for these systems a planetary source at the 100\% level, bar
for Kepler-699 with a probability of 99.1\%.} MCMC model fit results are
available for these systems at the NEA (Thompson {\em et al}, 2018;
Hoffman \& Rowe, 2017), which use quadratic limb darkening parameters
taken from Claret \& Bloemen (2011) and so not included as optimisable
parameters.  {  This paper's model attempts to fit limb
darkening.}}

\item Check these five systems with a second algorithm, {\em Winfitter}
(Rhodes \& Budding, 2014), which uses a Radau model that considers
ellipticity and reflection (Kopal, 1959). The program performs
optimisation by a modified Levenberg-Marquardt technique and estimates
error with Hessian inverse matrix (Budding~{\em et al.}, 2016a, b). 

\item Analyse possible sources of errors and suggest future improvements.
\end{itemize}
This project is similar to that described by Ji {\em et al} (2017), with
this paper being a partial extension of the work described by Ji, where
further background may be found.

\section{Simple Model}
The model was simple, essentially equivalent to that of Mandel \& Agol
(2002), {  using} the following parameters: the planet's
orbital radius $a$, the stellar radius $r_s$, the planet's radius $r_p$,
orbital inclination $i$, and {  initially} the linear limb
darkening coefficient $u$. Additional parameters $U$ and
$\textit{offset}$ were included to adjust the reference points of {  the} flux
axis and phase axis respectively.  {  An assumption of
Gaussian noise was included into the model, allowing a fit estimate to
be made (`sigma' in the following discussions).} The adoption of
circular orbits is a limitation of this model, {  along with
the use of linear limb darkening and the `small planet' approximation
(see, e.g., Nutzman {\em et al.}, 2009).  Inclination follows the usual
convention adopted by eclipsing binary light curve models, e.g.
$90\degree$ when in our line of sight.}

{  We started with linear limb darkening coefficients, given
the discussion of Budding {\em et al.} (2016) on the complexity of limb
darkening models and the information content of the modelled data. Other
authors such as Kipping (2010) and Csizmadia {\em et al.} (2013) note
the difficulty of extracting limb darkening coefficients from light
curves.  Nevertheless, following the initial linear term MCMC fits we
extended the model to quadratic limb darkening to see if we could obtain
co-efficients, and whether these were in line with Claret \& Bloemen
(2011).
}

 
\begin{table*}[htpb]
\begin{center}
\begin{tabular}{lcccc|cccc}
\hline 
             &  \multicolumn{4}{c}{Current Study} &   \multicolumn{4}{c}{{  Esteves {\em et al.}}} \\
System & ${r_{p}}/{r_{s}}$ & ${a}/{r_{s}}$ & $ i $ & $ u$ &  ${r_{p}}/{r_{s}}$ & ${a}/{r_{s}}$ & $ i $ & $ u$ \\
\hline\hline
Kepler-1b & 0.1277 & 7.8629 & 83.8115 & 0.665  & 
	$0.12539_{(-0.00035)}^{(+0.00049)}$ & $7.903_{(-0.016)}^{(+0.019)}$ & $83.872_{(-0.018)}^{(+0.020)}$ & 0.598 \\
Kepler-5b &  0.0795 & 6.1844 & 87.1027 & 0.443  &
	 $0.079965_{(-0.000071)}^{(+0.000087)}$ & $6.450_{(-0.025)}^{(+0.021)}$ & $89.14_{(-0.32)}^{(+0.44)}$ & 0.561\\
Kepler-8b & 0.0957 & 6.6869 & 83.6664 & 0.529  &
	$0.095751_{(-0.00023)}^{(+0.00019)}$ & $6.854_{(-0.017)}^{(+0.018)}$ & $83.978_{(-0.033)}^{(+0.036)}$ & 0.567	\\
Kepler-12b & 0.1191 & 7.7614 & 87.5911 & 0.486 &
	$0.118867_{(-0.000094)}^{(+0.000085)}$ & $8.019_{(-0.013)}^{(+0.014)}$ & $88.796_{(-0.074)}^{(+0.088)}$ & 0.589 \\
 \hline
\end{tabular}
\caption{\label{tab:lm_results} Parameter values from LM fits. The NEA
results are from Esteves {\em al.} (2015). {  The LM
methodology only provided point estimates and not `uncertainties',
explaining the need for a following step to be a method such as
boot-strapping or MCMC to better understand the accuracy of the
estimates.  The LM results do not overlap with the confidence intervals
of Esteves {\em et al.}, although without uncertainties for this paper's results only a
conclusion of general agreement between the two methods can be made.}}
\end{center}
\end{table*}


\section{Levenberg-Marquadt (LM) Fits}

The LM algorithm can be seen as a combination of the steepest gradient
algorithm and the Newton algorithm (Li {\em et al.}, 2017), providing
point estimates. Short cadence data from Quarter 1 were downloaded for
Kepler-1b, 5b, and 8b from the NEA website and folded using the given
(NEA) system periods. Quarter 2 short cadence data were used for
Kepler-12b.  Results for LM fits assuming only linear limb darkening are
given in table~\ref{tab:lm_results}, and compared with results from the
NEA. These are all similar (as well as to other studies such as Ji {\em
et al.}, 2017, and Budding {\em et al.}, 2016a,b), lending confidence to
our procedures.  { The LM results were used as starting
parameterisations for the subsequent Monte Carlo modelling.
The scipy.optimize.leastsq method was used for these optimisations
(Jones {\em et al}, 2001).
}


\section{MCMC}
In probability theory, a Markov chain is a sequence of random variables
$\theta^1, \theta^2, ...,$ {  in} which, for any $t$, the
distribution of $\theta^t$, even given all previous $\theta$'s, depends
only on the most recent value, $\theta^{t-1}$. Markov chain simulation
is a general method that draws values of $\theta$ from approximate
distributions, and then improves the draws at each subsequent step to
better approximate the target posterior distribution, $p(\theta |y)$
{  (where $y$ is the dependent variable)}. The sampling is
done sequentially, such that the sampled draws form a Markov chain
(Gelman~{\em et al.}, 2009). We used the Hamiltonian Monte Carlo (HMC)
algorithm, which suppresses the local random walk behaviour of the
classic Metropolis-Hasting algorithm, allowing faster exploration of the
target distribution. The algorithm was implemented using the
$\textit{pystan}$ package in Python\footnote{Stan Development Team
(2017). PyStan: the Python interface to Stan, Version 2.16.0.0.
\url{http://mc-stan.org}}. 


\begin{table*}[htpb]
\begin{center}
\begin{tabular}{lccccccc}
\hline 
 System & p ($r_{p}/r_{s} $) & or ($r_{s}/a$) & u & cos(i) & $\sigma$ (x  $10^{-6}$)  & T\\
\hline\hline
 Kepler-1b       & 0.1275 $\pm$ 0.0004 & 0.128 $\pm$ 0.003 & 0.636 $\pm$ 0.003 & 0.1084 $\pm$ 0.0005 & 42 $\pm$ 4 & 100\\
 Kepler-5b       & 0.0794 $\pm$ 0.0003 & 0.157 $\pm$ 0.002 & 0.431 $\pm$ 0.013 & 0.0301 $\pm$ 0.0129 & 112 $\pm$ 9 & 80 \\
 Kepler-8b       & 0.0958 $\pm$ 0.0018 & 0.147 $\pm$ 0.007 & 0.614 $\pm$ 0.085 & 0.1062 $\pm$ 0.0011 & 910  $\pm$  40 & 200\\
 Kepler-12b     & 0.1191 $\pm$ 0.0005 & 0.129 $\pm$ 0.002 & 0.486 $\pm$ 0.015 & 0.0417 $\pm$ 0.0060 & 160 $\pm $ 30 &  20 \\
 Kepler-77b     & 0.0994 $\pm$ 0.0004 & 0.107 $\pm$ 0.008 & 0.575 $\pm$ 0.017 & 0.0471 $\pm$ 0.0021 & 196 $\pm$  10 & 200 \\
 Kepler-428b    & 0.1467 $\pm$ 0.0002 & 0.119 $\pm$ 0.001 & 0.717 $\pm$ 0.011 & 0.0843 $\pm$ 0.0006 & 42 $\pm$ 8 & 150 \\
 Kepler-491b   & 0.0818 $\pm$ 0.0003 & 0.094 $\pm$ 0.001 & 0.599 $\pm$ 0.010 & 0.0489 $\pm$ 0.0022 & 86 $\pm$ 6 & 100\\
 Kepler-699b  & 0.1644 $\pm$ 0.0005 & 0.024 $\pm$ 0.001 & 0.722 $\pm$ 0.067 & 0.0192 $\pm$ 0.0002 & 61 $\pm$ 2 & 80 \\
 Kepler-706b   & 0.1423 $\pm$ 0.0004 & 0.016 $\pm$ 0.001 & 0.649 $\pm$ 0.011 & 0.0092 $\pm$ 0.0002 &  40 $\pm$ 13 & 80 \\
 Kepler-730b   & 0.0850 $\pm$ 0.0003 & 0.116 $\pm$ 0.001 & 0.618 $\pm$ 0.018 & 0.0871 $\pm$ 0.0018 & 52 $\pm$ 6 & 200 \\
\hline
\end{tabular}
\caption{\label{tab:mcmc_results} Parameter values from the linear limb darkening MCMC fits.  One standard deviation `errors' are given (as $\sigma$).
`p' is the planetary radius $r_{p}$ divided by the stellar radius $r_{s}$ , `or' the 
stellar radius $r_{s}$ divided by $a$ (the orbital semi-major axis),  `u' the linear limb darkening, 
and `cos(i)' the cosine of the orbital inclination. `T' is the number of steps, in thousands, taken following burn-in periods.  The chains were not thinned, given the ACF results discussed in the paper body. Howarth (2011) gives 0.5364 as the linear limb darkening
coefficient for Kepler-5 and 0.5850 for Kepler-8.  The second value is close to this paper's, but the first
is well outside reasonable `errors' based on the derived standard deviations.}
\end{center}
\end{table*}


\begin{table*}[htpb]
\begin{center}
\begin{tabular}{lccccc}
\hline 
 System & p ($r_{p}/r_{s} $) & or ($r_{s}/a$) & u & cos(i) & $\sigma$ (x  $10^{-6}$) \\
\hline\hline
 Kepler-5b & 0.0790 $\pm$ 0.0021 & 0.1592 $\pm$ 0.0177 & 0.477 $\pm$ 0.034 & 0.0393 $\pm$ 0.0005 & 64.5 \\
 Kepler-77b & 0.0994 $\pm$ 0.0016 & 0.1019 $\pm$ 0.0039 & 0.516 $\pm$ 0.038 & 0.0333 $\pm$ 0.0003 & 88.3 \\
 Kepler-428b & 0.1465 $\pm$ 0.0002 & 0.1170 $\pm$ 0.0011 & 0.683 $\pm$ 0.011 & 0.0810 $\pm$ 0.0006 & 41.8 \\
 Kepler-491b   & 0.0816 $\pm$ 0.0012 & 0.0861 $\pm$ 0.0030 & 0.594 $\pm$ 0.029 & 0.0311 $\pm$ 0.0004 & 50.7 \\
 Kepler-699b  & 0.1646 $\pm$ 0.0004 & 0.0237 $\pm$ 0.0001 & 0.644 $\pm$ 0.137 & 0.0187 $\pm$ 0.0002 & 62.4 \\
 Kepler-706b   & 0.1424 $\pm$ 0.0003 & 0.0162 $\pm$ 0.0001 & 0.627 $\pm$ 0.012 & 0.0085 $\pm$ 0.0002 &  43.2 \\
 Kepler-730b   & 0.0846 $\pm$ 0.0003 & 0.1117 $\pm$ 0.0007 & 0.615 $\pm$ 0.017 & 0.0805 $\pm$ 0.0011 & 51.6 \\
\hline
\end{tabular}
\caption{\label{tab:winfitter_results} Parameter values from {\em WinFitter} fits. Labels are as described at Table 2. }
\end{center}
\end{table*}


\begin{table*}[htpb]
\begin{center}
\begin{tabular}{lcccc}
\hline 
Cadence &  ${r_{p}}/{r_{s}}$ & ${a}/{r_{s}}$ & $ i (deg) $ & $ u$ \\
\hline\hline
Short & [0.0878, 0.0912] & [0.0878, 0.0912] &[87.331, 89.156] & [0.565, 0.650]  \\
Long &	[0.0880, 0.0911] & [0.1306, 0.1462] & [82.849, 84.043] & [0.399, 0.714] \\
 \hline
\end{tabular}
\caption{\label{tab:491_results} Comparison of optimal parameters for long and short cadence data for Kepler-491b}
\end{center}
\end{table*}

\begin{figure*}[htp]
\centering
\includegraphics[width=1.0\textwidth]{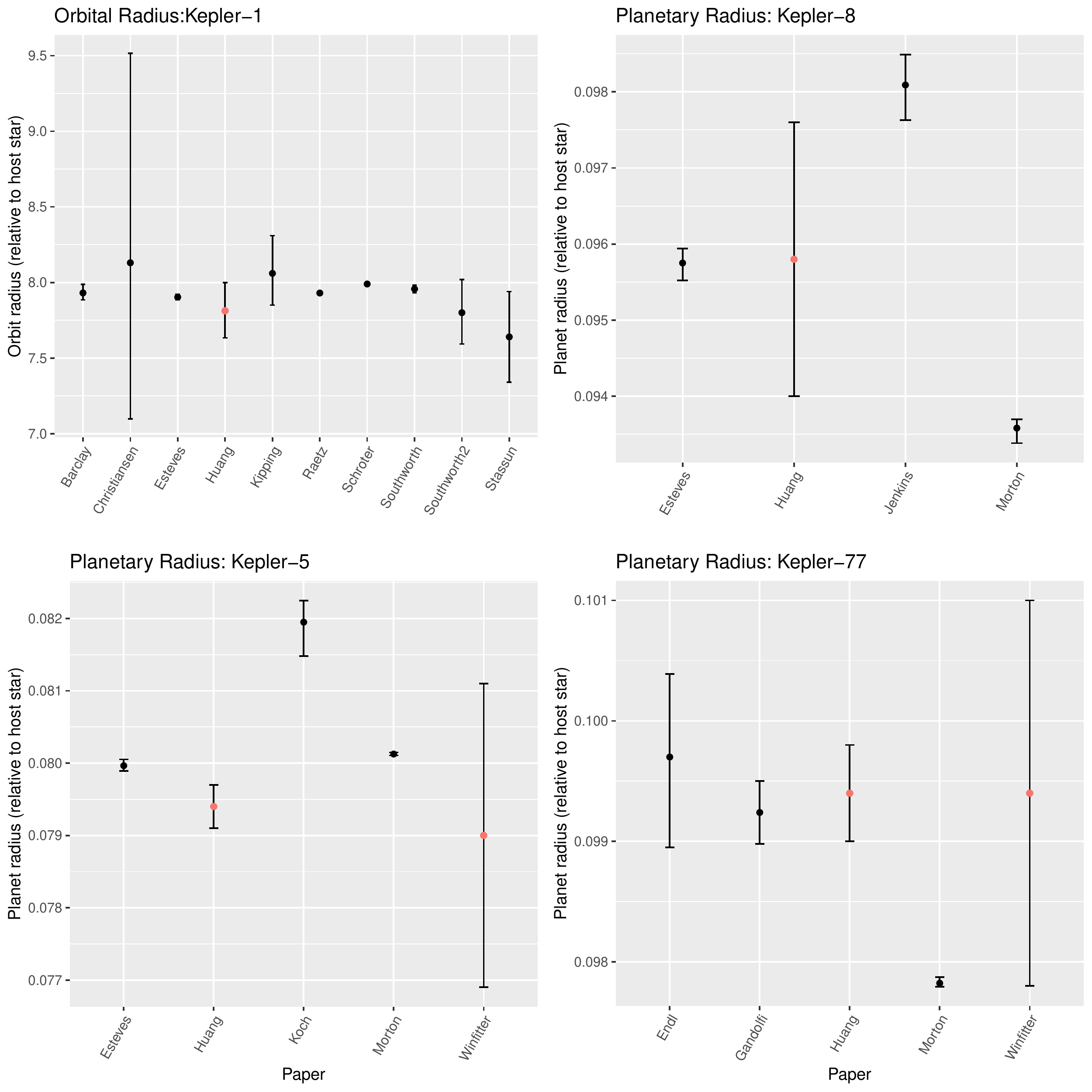}
\caption{\label{fig:comparison} Sample comparisons of parameters derived in this study (using MCMC as described
in the body) against the literature.  Papers are referred to by their primary author: 
Barclay {\em et al} (2012),
Christiansen {\em et al} (2011), 
Endl {\em et al.} (2014),
Esteves {\em et al} (2015),
Gandolfi {\em et al.} (2013)
Holczer {\em et al.} (2016),
Holman {\em et al.} (2007),
Kipping \& Bakos (2011),
Koch {\em et al.} (2010), 
Morton {\em et al.} (2016),  
Raetz ({\em et al.} (2014),
Schroter {\em et al} (2012),
Torres {\em et al.} (2008), and
Turner {\em et al.} (2016). 
{  Huang (this paper)} used {  68\% confidence intervals} from their MCMC fits as estimates of uncertainty,
{color{blue} as does {\em Winfitter} (based on examination of the error matrix, the inverse of the Hessian).} 
{  Errors for the other papers are as those papers reported them.}
}
\end{figure*}

\begin{figure*}[htp]
{ 
\centering
\includegraphics[width=0.8\textwidth]{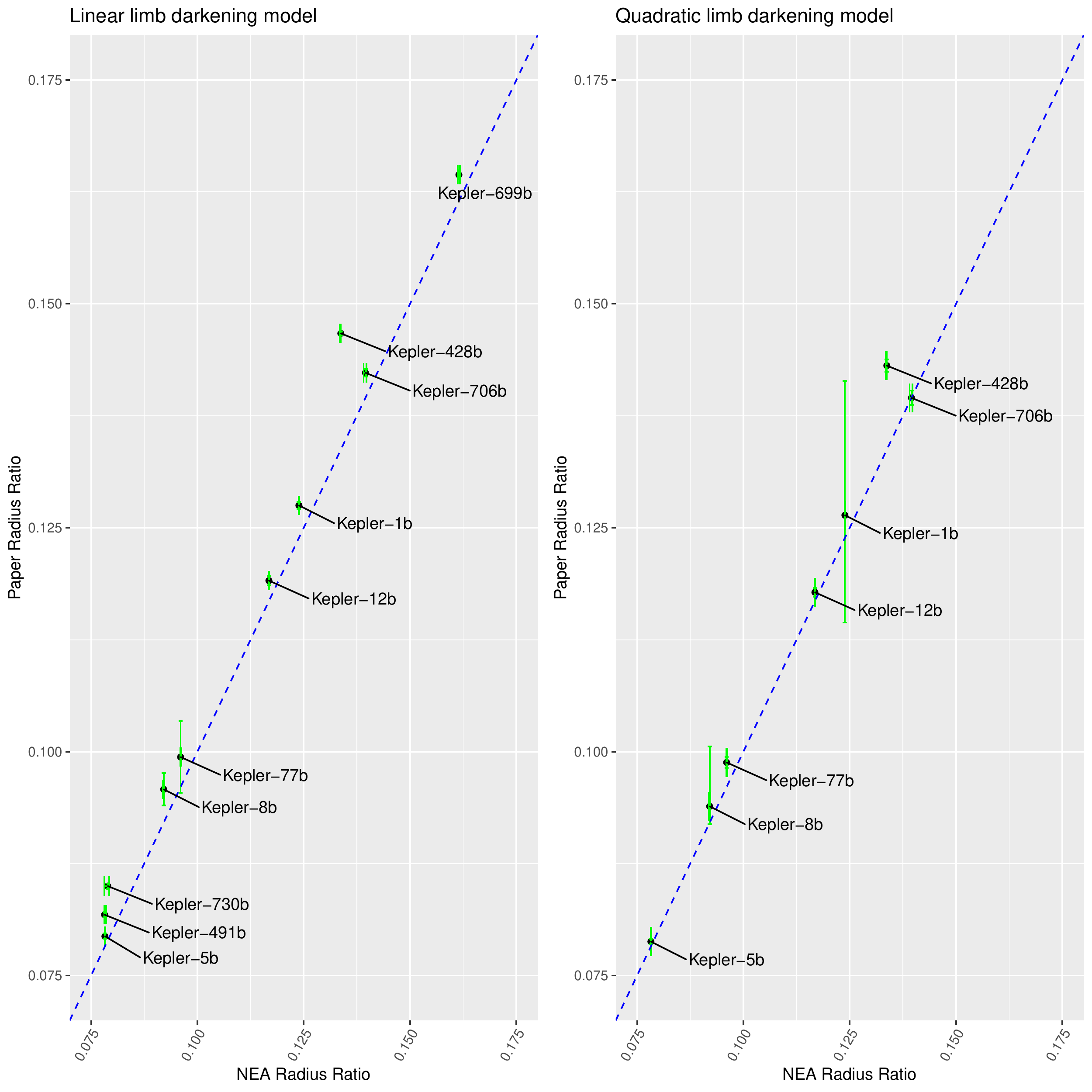}
\caption{\label{fig:comparison_nea}
Comparison of the planet to star radius from this paper (vertical axis) with those from the NEA MCMC fits (horizontal axis).  The chart on
the left is based in the linear limb darkening model, and on the right the quadratic limb darkening. The blue
dotted lines are those of perfect agreement, not the lines of best fit from linear regressions.}
}
 \end{figure*}

To reduce the impact of the starting values, we {  discarded} the first
half of each sequence before carrying out any analysis and inference.
This practice of discarding early iterations in Markov chain simulation
is referred to as {  discarding the }`warm-up'.

\begin{figure*}[htpb]
\centering
\includegraphics[width=1.0\textwidth]{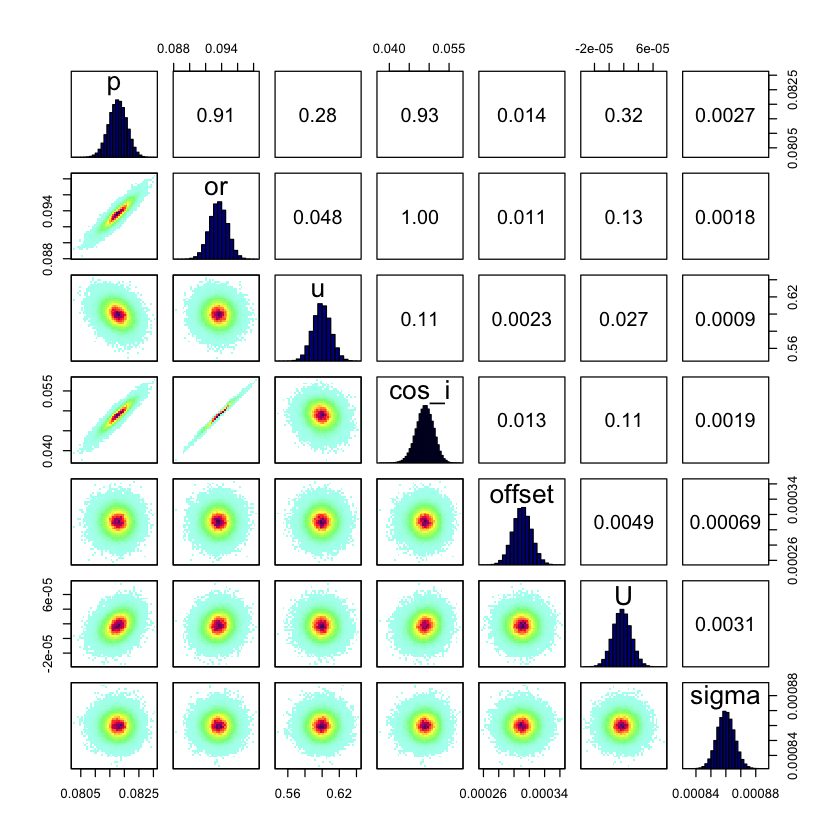}
\caption{\label{fig:491mcmc} Correlation plot for Kepler 491 short
cadence dataset, based on MCMC {  modelling} using 4 chains.
The density plots (in the lower left of the diagram) plot these 100,000
points for each parameter of the light curve model. `p' is the planetary
radius $r_{p}$ divided by the stellar radius $r_{s}$ , `or' the stellar
radius $r_{s}$ divided by $a$ (the orbital semi-major axis), `u' the
linear limb darkening, `cos\_i' the cosine of the orbital inclination,
`offset' the phase offset of the folded light curve, `U' the overall
flux adjustment, and `sigma' the Gaussian noise in the binned data (1
standard deviation).  The histograms along the diagonal show the
``error'' of the derived parameters (as well as the maximum likelihood),
while the numbers in the upper right of the diagram are the correlation
{  coefficients} between pairs of the parameters.
}
\end{figure*}

\begin{figure}[htpb]
\centering
\includegraphics[width=0.40\textwidth]{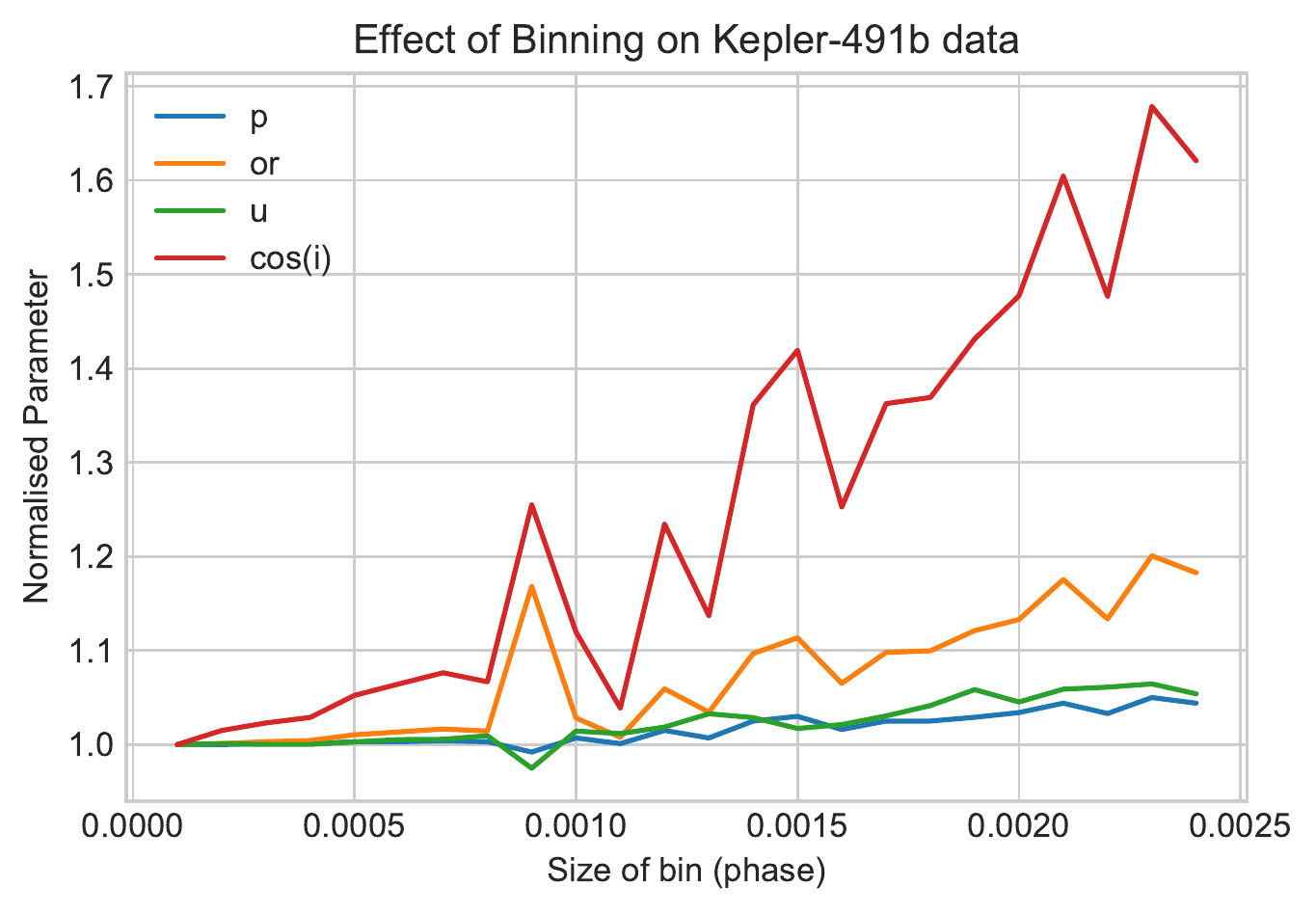}
\caption[\label{fig:491} Binning test using Kepler 491 short cadence
dataset.] {\label{fig:491} Binning test using Kepler 491 short cadence
dataset.} 
\end{figure}


\begin{table*}[htpb]
{ 
\begin{center}
\begin{tabular}{lllllllllll}
\hline 
System & p ($r_{p}/r_{s} $) & or ($r_{s}/a$) & $u_{1}$ & $u_{2}$ & cos(i) & $\sigma$ (x  $10^{-6}$)  \\
\hline
Kepler-1b    & $0.1264_{-0.0012}^{+0.0015}$ & $0.1296_{-0.0011}^{+0.0016}$ & $0.33_{-0.22}^{+0.17}$ &  $0.40_{-0.21}^{+0.25}$ &
	$0.110_{-0.01}^{+0.02}$ & 44 \\
Kepler-5b    & 0.0788 $\pm$ 0.0003 &  0.156 $\pm$ 0.014    & 0.31 $\pm$ 0.04 & 0.26 $\pm$ 0.09 & 0.019 $\pm$ 0.011 & 107 \\
Kepler-8b    & $0.0939_{-0.0020}^{+0.0067}$ & $0.1470_{-0.0065}^{+0.0156}$ & $0.30_{-0.16}^{+0.54}$ &  $0.77_{-0.25}^{+0.20}$ &
	$0.104_{-0.01}^{+0.06}$ & 919 \\
Kepler-12b  & 0.1178 $\pm$ 0.0005 &  0.128 $\pm$ 0.001    & 0.38 $\pm$ 0.03 & 0.25 $\pm$ 0.07 & 0.037 $\pm $ 0.004 &  103 \\
Kepler-77b  & 0.0988 $\pm$ 0.0006 &  0.106 $\pm$ 0.001    & 0.47 $\pm$ 0.06 & 0.19 $\pm$ 0.11 & 0.045 $\pm$ 0.002 & 197 \\
Kepler-428b& 0.1431 $\pm$ 0.0007 &  0.122 $\pm$ 0.001    & 0.22 $\pm$ 0.09 & 0.80 $\pm$ 0.15 & 0.0871 $\pm$ 0.0009 & 135 \\
Kepler-706b& 0.1395 $\pm$ 0.0008 &  0.0166 $\pm$ 0.0001& 0.34 $\pm$ 0.08 & 0.61 $\pm$ 0.14 & 0.0091 $\pm$ 0.0003 & 42 \\
\hline
\end{tabular}
\caption{\label{tab:quadratic_results}
Parameter values from the quadratic limb darkening MCMC fits. 
Post-warmup chains were 200,000 steps.  $u_1$ is the first coefficient in quadratic limb darkening,
and $u_2$ the second. The errors for Kepler-1b and 8b
were not Gaussian, so median values are reported with differences to
upper and lower quartiles to give indication of spread.  
Kipping (2010) reported $u_1$ values of 0.38 and 0.14 for
his fits to {\em Kepler} short and long cadence Kepler-1b data
respectively, and $u_2$ values of 0.20 and 0.46 for the two data sets. 
Unfortunately the large derived errors in this study (for short cadence
data) make a comparison meaningless, although emphasizing his comment on
the difficulty of deriving limb darkening coefficients from these data.
Kepler-699b failed to stabilise at a physical solution, pushing both
$u_1$ and $u_2$ to zero and p towards one. Kepler-491b and 730b also failed, indicating 
full indeterminacy for $u_2$.
}
\end{center}
}
\end{table*}

\begin{figure*}[htp]
{ 
\centering
\includegraphics[width=0.8\textwidth]{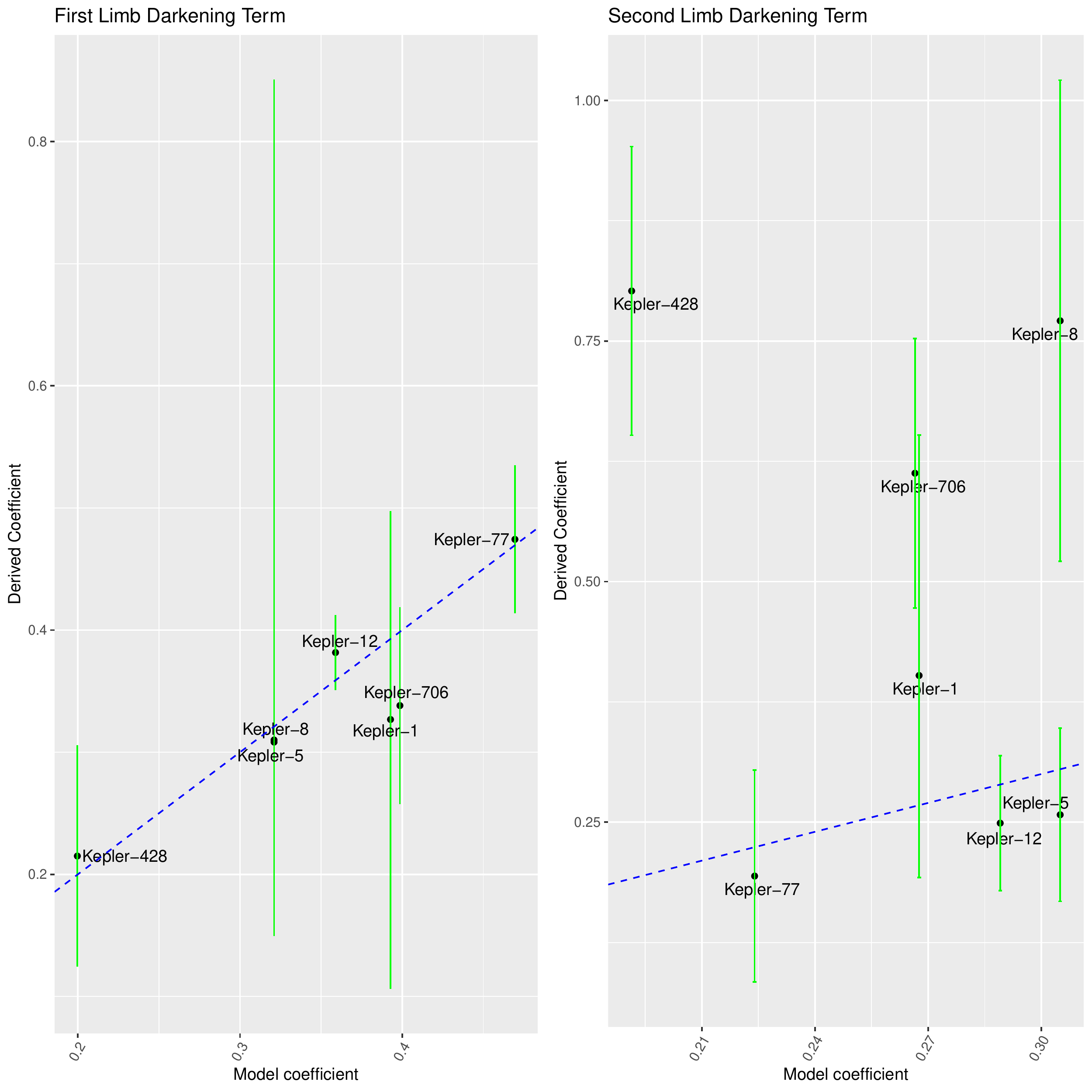}
\caption{\label{fig:comparison_ld}
Comparison between the quadratic limb darkening terms derived from the
MCMC model fits and those of Claret \& Bloemen (2011).  The latter are
used as fixed input by Thompson {\em et al} (2018) and Hoffman \& Rowe
(2017) in their MCMC fits.  The blue coloured dotted lines are those of
perfect agreement between the data sets (slope 1, intercept 0), showing
that the first term is in better agreement better the two data sets than
the second term.  
}
}
 \end{figure*}

\begin{figure}[htpb]
\centering
\includegraphics[width=0.45\textwidth]{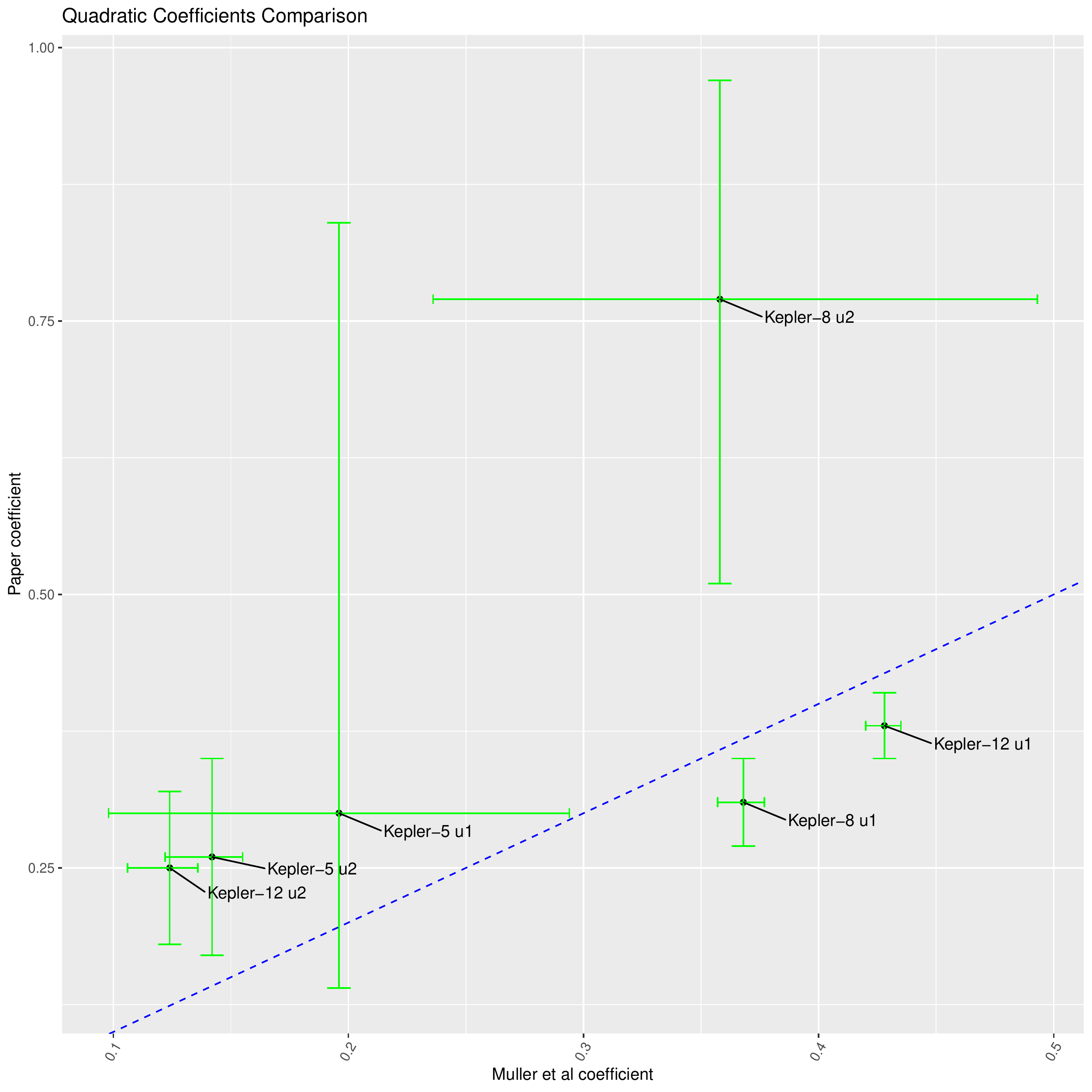}
{ 
\caption[\label{fig:muller} Comparison of Muller {\em et al} quadratic
limb darkening coefficients] {\label{fig:muller} Comparison of Muller
{\em et al} (2013) quadratic limb darkening coefficients for systems in
common with current paper.  Muller {\em et al.} modelled short cadence
{\em Kepler} data. The estimates do not fall (within errors) on the
marked line  (slope 1, intercept 0), which would indicate agreement.
`u1' is the first quadratic term, and `u2' the second.}
}
\end{figure}


\subsection{MCMC Results}

The data for the MCMC test systems were the same as used in the LM fits.
Table~\ref{tab:mcmc_results}  {  summarises} the results for
the systems, {  where the model included linear limb
darkening}. Figure~\ref{fig:comparison} shows a sample of comparisons
for system parameters from this {  table} compared with
literature results, showing a general agreement that we took as
encouraging. It also clearly demonstrates the frequently different
estimates for uncertainties.  The left hand plot of
Figure~\ref{fig:comparison_nea} plots the ratio of the planet to stellar
radii for the NEA MCMC fits against those from this paper's linear limb
darkening models.  A linear regression indicates a 2.1\% difference in
the estimates (this paper giving systematically larger radii), although
with a 3.9\% standard error in this estimate.  The intercept was 0.002
$\pm$ 0.004.  These results should be contrasted with those in
Figure~\ref{fig:comparison}, which shows similar magnitude range in
scatter across studies.

We therefore moved to the step of modelling systems without published
MCMC analyses {  including limb darkening as free
parameters}. 1600 days of Kepler data starting BJD 2454833 were folded
using the NEA published period and {  modelling} was carried
out  for systems Kepler-428b, 699-b, and 706b.  Kepler-491b data covered
160 days from BJD 2455333.

Kepler data are available in two cadences, short cadence and long
cadence. Each cadence is composed of multiple 6.02-s exposures with
associated 0.52-s readout times (Gilliland {\em et al.}, 2011). There is
a longer time interval between observations for long and short cadence
data which means there are 30 times more short cadence  than long
cadence data points in a quarter.\footnote{See
https://keplergo.arc.nasa.gov/DataAnalysisProducts.shtml for further
technical information on the {\em Kepler} mission and its imagery.}

Long cadence data were used for Kepler-428b, 699b, 706b, and 730b. Short
cadence data were used for Kepler-491b, as these were available for this
system, but not for the other three. The chains all converged well. The
trace plots exhibited rapid up-and-down variations with no long term
trends, indicating good mixing and that the Markov chains explored well
the posterior distributions. The ACF (auto-correlation function) plots
all decayed rapidly for each {  modelled} parameter.
$\hat{R}$ statistics were all close to 1, suggesting good convergence.
{  Results are given in Table~\ref{tab:mcmc_results}. As
discussed in Brooks \& Gelman (1998), if $\hat{R}$ is less than 1.2 then
the chains are approximately converged.} Figure~\ref{fig:491mcmc} is an
example correlation plot from this {  modelling}, and
representative for the {  modelled} systems.

To confirm the results from the simple model used in this paper, we used
{\em Winfitter} (see, e.g., Budding {\em et al.}, 2016) and its more
detailed fitting function to model the same data.
Table~\ref{tab:winfitter_results} presents the results for key
parameters, which are in good agreement with the simple model estimates,
lending confidence in them.

Kepler-491b had both long and short cadence data available, allowing us
to explore whether the derived parameters were affected by the implicit
data binning.   Table~\ref{tab:491_results} {  summarises}
the interval estimates given by the HMC algorithm for short and long
cadence datasets. When long cadence data are used to fit our light curve
model, the estimates for the parameters $\frac{r_s}{a}$ and
$\frac{r_p}{r_s}$ are high while the estimate for the parameter $i$ is
considerably reduced. This suggests that the use of long cadence dataset
may systematically overestimate the radii of a planet and host star,
while underestimating the planetary inclination.

We further investigated this binning effect using the Kepler 491b short
cadence dataset. Figure~\ref{fig:491} shows the percentage change in the
point estimates of the transit parameters given by the HMC algorithm for
various bin widths, relative to the point estimates derived from the
first bin width (0.0001 phase bin). We can see that when the bin width
increases, the derived radii $r_p$ and $r_s$ become larger while the
inclination $i$ becomes smaller (derived parameters $\frac{r_p}{r_s}$,
$\frac{r_s}{a}$, $\cos(i)$ increase as bin width increases).
{  This point was made by Kipping (2010), who recommended to
first compute the transit model at a finer time sampling, and then
integrate the ``supersampled" model over the observed integration time
before comparing it to the data.  The present analysis has independently
confirmed the underlying point, but rather than rework the model's
approximations for different sampling bins we call attention to the
observable effects of finite sampling in the residuals, since this might
be associated with some physical effect, such as abnormal
limb-darkening.}

Murphy (2012) commented that `the short cadence data are almost always
better than the long cadence data'. Our analysis confirms this
observation, at least for the systems we modelled.

We have only tested the impact of long cadence data by one system. For
other planetary systems having similar transit times and orbital periods
as Kepler-491b, we can expect a similar impact of the cadence value on
the parameters. For planetary systems that have longer transit
durations, the use of long cadence data may not be of such importance.

{  

Finally, we returned to the matter of including quadratic
limb darkening (see Table~\ref{tab:quadratic_results}), including it into the MCMC fits. 
Convergence could not be obtained for all systems (e.g., 491, 699, and 730), 
suggesting we were attempting to
extract `too much' information from the data.

The right hand plot of Figure~\ref{fig:comparison_nea}
compares the radii ratios for the systems with quadratic limb darkening fits in this
paper against the NEA estimates. 
The difference for the other systems
was within approximately one percent and within the combined error of
the estimates (but outside the formal errors for the radii given by the
linear limb darkening fits).   There is poor agreement between the models
of Claret \& Bloemen (2011) and the fit results for the $u_2$ limb
darkening values, but better for $u_1$ (see Figure
\ref{fig:comparison_ld}).

Howarth (2011) presented  simulations where use of the linear law led to
systematic errors of up to 4\% in radius estimates, but negligible error
introduced when the quadratic law was used.  This study finds a mean
increase of 1.4 $\pm$ 0.8\% increase in the planetary radii ratios using
linear limb darkening to those using quadratic, but this difference was
driven by Kepler 428 and 706 with their larger differences. 

\section{{\em Winfitter}  Quadratic Limb Darkening}

We then attempted {\em Winfitter} fits including quadratic limb
darkening.  {\em Winfitter} evaluates  the $\chi^2$ Hessian (see, for
example, Bevington, 1969) in the vicinity of the derived minimum. 
Inspection of this matrix, and in particular its eigenvalues and
eigenvectors, gives insights into parameter determinancy and
interdependence.  The Hessian can be inverted to yield an error matrix. 
A positive definite matrix indicates a determinate, `unique' solution.
This makes it possible to determine which of the adjustable parameters
are likely to allow well-defined optimal values, as well as providing an
error range for the optimized parameters. Further information on how
{\em Winfitter} evaluates the information content of data may be found
in Banks~\& Budding (1990).  Further details on {\em Winfitter} and its
usage may be found in Budding~\& Najim (1980), Budding~\& Zeilik (1987),
and Budding~\& Demircan (2007).

Only the fits for Kepler-5 and Kepler-77 were positive definite when
quadratic limb darkening was included into the model, indicating that
the information content of the data was being exceeded.  Inclination
tended to be the variable `breaking', which would be inline with its
high correlation with radii (see, for example,
Figure~\ref{fig:491mcmc}). However {\em Winfitter} indicated large
errors for the limb darkening coefficients for these two `successful'
systems: $u_{1} = -0.01 \pm 0.26$ and $u_{2} = 0.30 \pm 0.20$ for
Kepler-5 $u_{1} = 0.30 \pm 0.46$ and $u_{2} = 0.13 \pm 0.32$ for
Kepler-77.  These large errors are symptomatic of near breakdown of
determinacy.  All fits with linear limb darkening were positive definite. 

We do not expect it too surprising that quadratic limb darkening
coefficients could not be reliably `solved' for the long integration
data sets with the current methods, nor that it would be challenging
with the short integration {\em Kepler} data. For example, previous MCMC
optimisations by Ji {\em et al} (2017) had met the same problems with
Kepler-1 and Southworth (2009) had commented that ``....the linear law
is adequate for most of the datasets studied in this work (particularly
those from longer wavelengths)'' in his study of exoplanet transit light
curves.

Putting to one side our information limit concerns, we also tried a two
stage fitting approach to see what the effect on radii of quadratic limb
darkening would be compared to linear.  We first fitted the ratio of the
radii, the limb darkening coefficients, the stellar radius, and the
inclination (as well as adjustments in flux and time), followed by
removing the radii from the parameters optimised in a second fit.  This
led to planetary radius estimates from the quadratic fits that were
systematically biased compared to those from the linear limb darkening
fits  (by $ 0.991 \pm  0.004$) and in stellar radii by $1.050 \pm  
0.018$. We do not put much weight on these ratios, given the small data
set. There was poor agreement between the derived limb darkening
parameters and those of Claret \& Bloemen (2011). The Pearson
correlation coefficients were $-0.44$ for $u_1$ and $-0.31$ for $u_2$.

Attempts to determine limb darkening coefficients are important, as
exoplanet transit light curves provide a `laboratory', hopefully
providing data that can be used to test and refine stellar models (see
also Csizmadia {\em et al}, 2013).  Higher signal to noise data would be
ideal for these tests, particularly given the correlation between limb
darkening and our key variable of interest, the exoplanet radius. 
}


\section{Conclusions}

{  Kepler-1b,  5b,  8b, 12b, 77b, } 428b, 491b,  699b, 706b, and 730b were
analyzed with two approaches: an MCMC one based on Mandel \& Agol (2002)
and the other {\em Winfitter} on Kopal (1959). {  
These systems do not show any significant level of tidal and rotational distortion, for instance, by visual inspection of the light curves.
The coefficients for
gravity-darkening and stellar reflectivity were not optimised in the
majority of the {\em Winfitter} fits, bar for Kepler-699, 706, and 730  where the
reflection coefficient was found to be zero ($\pm$ 0.0001, $\pm$ 0.0002,
and $\pm$  0.0001 respectively).\footnote{ We subsequently ran {\em Winfitter} fits including
these as free parameters into the optimisation.  The information limit was exceeded,
indicating invalid solutions.} Further details on how ellipticity is
included into the model may be found in Kopal (1959).} In general, for
these systems the {\em Winfitter} results are in good agreement with the parameter
values from the MCMC fits. However the fractional stellar radius
and inclination of the Kepler-491 system are different to that from the 
MCMC fits.  There is a relationship or correlation between the two
variables (see Fig~\ref{fig:491mcmc}), so while it is disappointing that
there is a difference in the estimates it can be understood.


We have three major conclusions from this study:

\begin{itemize}

\item{
We note difficulty in consistent error estimates across
methodologies, as given in the literature.  As shown in
Fig~\ref{fig:comparison}, there can be a wide range in error estimates
for the same system, from many multiples larger than the estimates from
the MCMC and {\em Winfitter} fits of this paper, to many multiples less.  
However it is encouraging that the point estimates are in reasonable
agreement. Regardless, care will need to be taken with meta-analyses
based on the literature to avoid over-interpretation, particularly for
systems with perhaps only one published solution.
}

\item{
{  It appears difficult to derive limb darkening coefficients
through fitting, noting that we should not have used the long cadence
data for such modelling without compensating for the integration times.
However, Csizmadia {\em et al.} (2013) recommend trying to fit these
parameters, noting that some authors have highly different observed limb
darkening coefficients from the theoretical predictions (e.g., Claret,
2009; Kipping \& Bakos, 2011, Barros {\em et al.}, 2009).  It is hoped
that similar MCMC fittings could provide useful data to test stellar
models (see e.g., Figure~\ref{fig:muller} for a comparison of common
systems in one such paper with the current paper) and the reliability of
such estimates.  As Muller {\em et al} (2013) note, the upcoming PLATO
and JWST missions, along with the current TESS mission, should provide
high signal to noise data that will allow deeper investigations of limb
darkening including any diversity with effective temperature etc.  Southworth
(2008, 2009, 2010, 2011, 2012) provides an interesting self-consistent survey
with careful consideration of errors, worthy of emulation.}
}

\item{
Consideration of integration periods will prove important for analysis
of some systems where observations have long integration times compared
to transit ingress or egress. Examples could include the TESS mission
for selected stars observed with 30-minute time sampling in the
full-frame images (FFIs, see Bouma {\em et al.}, 2017, for further
information on TESS). Such binning will, naturally, be increasingly
important for transits where the integration times are significant
fractions of the total transit time.  Flux measurements during the
ingress and egress periods will be `smeared out', leading to wider
estimates of the planetary radius, as we have seen in this study.  A
simple model, as we have used in this paper, will lead to systematic
biases in the derived parameters unless compensation is made for the
longer integration times. {  We intend to add this feature into later
work with {\em Winfitter} on long integration data}.
}

{ 
We did not explore thoroughly in this paper a detailed comparison of the
formal errors produced by the MCMC and {\em Winfitter} methodologies, as
believe this is worthy of a more detailed follow-up study.  We intend to
implement MCMC as the optimisation technique for the {\em Winfitter}
model, allowing deeper examination of the errors estimated by this
methodology. {\em Winfitter} is not only a more sophisticated model, for
instance including relevant proximity effects (such as radiative
interaction, tidal and rotational distortions) as well as orbital
ellipticity with the methodology to include simple modeling of
starspots.  It is much faster in computer runtimes than MCMC\footnote{
  For example, {\em Wifitter} runtimes on a single core
VMware Windows 10 emulation on an i-5 2012 MacBook Pro were of the order
of seconds, extending to inside a minute for high (approx. 500) numbers
of iterations.  STAN code on the same hardware and in the native OS were
of 10-12 hours per chain (of the size reported in this paper).  STAN currently does not support GPUs although
it does support multicore.  We will be looking into alternative
frameworks for GPU support, to further reduce computing time.
}, and if it can be shown
that its error estimates are comparable to those from MCMC, it could be
a useful tool for large scale studies across multiple exoplanet systems.
 A comparison of the formal errors for the systems in this paper is
inconclusive --- for instance, {\em Winfitter} tended to underestimate
(in comparison with the MCMC model) errors in $\cos{i}$, was generally
larger in $u$, and mixed in stellar and planetary radii.  We intend to
widen the data set, using short integration time data.
}

\end{itemize}


\section{Acknowledgements}

This research has made use of the NASA Exoplanet Archive, which is
operated by the California Institute of Technology, under contract with
the National Aeronautics and Space Administration via the Exoplanet
Exploration Program.  It is a pleasure to acknowledge additional help
and encouragement from the National University of Singapore (NUS),
particularly through Prof.\ Lim Tiong Wee of the Department of
Statistics and Applied Probability. This paper reports results from
an undergraduate student project at NUS, {  in that department.
{\em Winfitter} may be sourced from https://michaelrhodesbyu.weebly.com/astronomy.html.
We thank the referees of this paper for their helpful comments, which helped
improve the paper.}



\begin{thebibliography}{}
 
\bibitem[\protect\citeauthoryear{akeson}{2013}]{ake} {
Akeson, R.L., Chen, X., Ciardi, D., Crane, M., Good, J., Harbut, M., 
Jackson, E., Kane, S.R., Laity, A.C., Leifer, S., Lynn, M., McElroy, D. L., 
Papin, M.,  Plavchan, P.,  Ramirez, S.V.,  Rey, R., von Braun, K., 
Wittman, M. , Abajian, M.,  Ali, B., Beichman, C.,  Beekley, A., Berriman, G.B.,
Berukoff, S.,  Bryden, G.,  Chan, B.,  Groom, S.,  Lau, C.,  Payne, A. N., 
Regelson, M.,  Saucedo, M.,  Schmitz, M., Stauffer,  J.,
Wyatt, P., \& Zhang, A., PASP, 125, 989, 2013}

{ 
\bibitem[\protect\citeauthoryear{Banks}{1990}]{banks} 
Banks, T., \& Budding, E., 1990, Ap\&SS, 167, 221
}

\bibitem[\protect\citeauthoryear{Barclay}{2012}]{bar} Barclay, T.,  Huber, D., 
Rowe, J.F., Fortney, J.J., Morley, C.V., Quintana, E.V., Fabrcky, D.C., Barentsen, G., 
Bloemen, S., Christiansen, J.L., Demory, B-O, Fulton, B.J., Jenkins, J.M., Mullally, F., 
Ragozzine, D., Seader, S.E., Shporer, A., Tenebaum, P., \& Thompson, S.E.,  2012
Ap. J, 761(1),  53

{ 
\bibitem[\protect\citeauthoryear{Barrosi}{2012}]{barros} 
Barros, S. C. C., Pollacco, D. L., Gibson, N. P., {\em et al.},  2012, MNRAS, 419, 1248
}
 
 { 
\bibitem[\protect\citeauthoryear{Bevingtoni}{1969}]{Bevington} 
Bevington, P. R., 1969, {\em Data Reduction and Error Analysis for the 
Physical Sciences }, McGraw-Hill, New York
}
 
\bibitem[\protect\citeauthoryear{Borucki}{2003}]{bor} Borucki, W.J., et
al.\ (14 authors), 2003, in {\em Scientific Frontiers in Research on
Extrasolar Planets}, Eds.\ D.\ Deming and S.\ Seager, ASP Conf.\ Ser.\,
294, 427

\bibitem[\protect\citeauthoryear{}{2011}]{bo2} Borucki, W.J., et
al.\ (69 authors), 2011, ApJ, 736, 19

\bibitem[\protect\citeauthoryear{}{2017}]{bo23}
Bouma, L. G., Winn, Joshua N., Kosiarek, Jacobi, et al.\ 
2017, pre-print (arXiv170508891B)

{ 
 \bibitem[\protect\citeauthoryear{}{1998}]{brooks}
 Brooks, S., \& Gelman, A., 1997, J. Comput. Graph Stat., 7(4), 434
}

{ 
 \bibitem[\protect\citeauthoryear{}{1980}]{najim}
 Budding, E., \& Najim, N. N., 1980, Ap \& SS, 72, 369.
 }
 
 { 
 \bibitem[\protect\citeauthoryear{}{2007}]{Demircan}
 Budding, E., \& Demircan, O., 2007, {\em Introduction of Astronomical Photometry},
 Cambridge University Press, Cambridge.
 }
 
 { 
 \bibitem[\protect\citeauthoryear{}{1987}]{Zeilik}
 Budding, E., \& Zeilik, M., 1987, Ap. J., 319, 827
 }
 
 { 
 \bibitem[\protect\citeauthoryear{Budding}{2016a}]{bu3} Budding, E.,
P\"{u}sk\"{u}ll\"{u}, \c{C}., Rhodes, M.D., Demircan, O., \& Erdem, A.,
2016a, Ap{\&}SS, 361, 17

\bibitem[\protect\citeauthoryear{Budding}{2016b}]{bu4} Budding, E.,
Rhodes, M.D., P\"{u}sk\"{u}ll\"{u}, \c{C}., Ji, Y., Erdem, A., \& Banks,
T.,  2016b, Ap{\&}SS, 361, 346
}
 
{ 
\bibitem[\protect\citeauthoryear{}{2009}]{claret1}
Claret, A., 2009,  A\&A, 506, 1335
}

{ 
\bibitem[\protect\citeauthoryear{}{2011}]{claret}
Claret, A., \& Bloemen, S., 2011, A\&A, 529, 75
}

{ 
\bibitem[\protect\citeauthoryear{Csizmadia}{2013}]{ecsizmadia}
{
Csizmadia, Sz., Pasternacki, Th., Dreyer, C., Cabrera, J., Erikson, A., \& Rauer, H.,
2013, A\&A, 549, A9
}
}

\bibitem[\protect\citeauthoryear{Budding}{2007}]{bu2} Budding, E., \&
Demircan, O., 2007, {\em Introduction to Astronomical Photometry}, CUP

\bibitem[\protect\citeauthoryear{Budding}{2016a}]{bu3} Budding, E.,
P\"{u}sk\"{u}ll\"{u}, \c{C}., Rhodes, M.D., Demircan, O., \& Erdem, A.,
2016a, Ap{\&}SS, 361, 17

\bibitem[\protect\citeauthoryear{Budding}{2016b}]{bu4} Budding, E.,
Rhodes, M.D., P\"{u}sk\"{u}ll\"{u}, \c{C}., Ji, Y., Erdem, A., \& Banks,
T.,  2016b, Ap{\&}SS, 361, 346

\bibitem[\protect\citeauthoryear{Charbonneau}{1999}]{ch2} Charbonneau,
D., Noyes, R.W., Korzennik, S.G., Nisenson, P., Jha, S., Vogt, S.S., \&
Kibrick, R.I., 1999, ApJ, 527, 445

\bibitem[\protect\citeauthoryear{Christiansen}{2011}]{christiansen}
Christiansen, J.L., Ballard, S.,  Charbonneau, D., Deming, D., Holman, M.J.,
Madhusudhan, N., Seager, S., Wellnitz, D.D., Barry, R.K., Livengood, T.A.,
Hewagama, R., Hampton, D.L., Lisse, C.M., \& A'Hearn, M.F., 2011,
Ap. J., 726, 94

\bibitem[\protect\citeauthoryear{Endl}{2014}]{endl} 
Endl, M., Caldwell, D. A., Barclay, T., {\em et al.}, 2014,
Ap. J., 795, 151

\bibitem[\protect\citeauthoryear{Esteves}{2015}]{esteves} 
Esteves, L.J., De Mooij, E.,J.W., \& Jayawardhana, R., 2015,
{\em Ap. J}, 804(2), 28

\bibitem[\protect\citeauthoryear{gandolfi}{2015}]{gandolfi} 
Gandolfi, D., Parviainen, H., Fridlund, M., {\em et al.}, 2013,
A \& A, 557, 74

\bibitem[\protect\citeauthoryear{gelman}{2009}]{gelman} 
Gelman, A., Carlin, J.B., Stern, H.S., \& Rubin, D.B.,
Bayesian Data Analysis --- Second Edition, 2009, Chapman \& Hall/CRC.

\bibitem[\protect\citeauthoryear{Giliand}{2011}]{gilliand}
Gilliland, R. L., Chaplin, W. J., Dunham, E. W., {\em et al.}, 2011,
{ApJS}, 197, 6.

{ 
\bibitem[\protect\citeauthoryear{Hoffman}{2017}]{hoffman}
Hoffman, K. L., \&  Rowe, J. F., 2017,
{\em Uniform Modeling of KOIs: MCMC Notes for Data Release 25},
KSCI-19113-001 (https://exoplanetarchive.ipac.caltech.edu/\\docs/KSCI-19113-001.pdf)
}

\bibitem[\protect\citeauthoryear{Holczer}{2016}]{holczer}
Holczer, T., Mazeh, T., Nachmani, G., Jontof-Hutter, D.,
Ford, E. B., Fabrycky, D., Ragozzine, D., Kane, M., \& Steffen, J. H.,
2016, Ap. J SS, 225, 9

\bibitem[\protect\citeauthoryear{Holman}{2007}]{holman}
Holman, M.J., Winn, J.N., Latham, D.W., O'Donovan, F. T., 
Charbonneau, D., Torres, G., Sozzetti, A., Fernadez, J., \&
Everett, M.E., 2007, Ap. J., 664, 1185.

\bibitem[\protect\citeauthoryear{Howarth}{2011}]{Howarth}
Howarth, I.D., 2011, MNRAS, 418, 1165


\bibitem[\protect\citeauthoryear{Ji}{2017}]{jiyi}
Ji, Y., Banks, T., Budding, E., \& Rhodes, M.D., 2017, Ap{\&}SS, 362, 12

\bibitem[\protect\citeauthoryear{Jones}{2001}]{jones}
Jones, E., Oliphant, E., Peterson, P., et al., 2001,
\textit{SciPy: Open Source Scientific Tools for Python}.
Accessed Dec 12, 2017 from \url{http://www.scipy.org/}

{ 
\bibitem[\protect\citeauthoryear{Kipping}{2010}]{kipping1}
Kipping, D., 2010, MNRAS, 408 (3), 1758
}

\bibitem[\protect\citeauthoryear{Kipping}{2017}]{kipping}
Kipping, D., \& Bakos, G., 2011, Ap. J., 733, 36

\bibitem[\protect\citeauthoryear{Koch}{2010}]{koch}
Koch, D. G., Borucki, W. J., Rowe, J. F., {\em et al}, 2010, Ap. J Letters, 713, 131

\bibitem[\protect\citeauthoryear{Kopal}{1959}]{kopal}
Kopal, Z., 1959, {\em Close Binary Systems}, London, Chapman \& Hall.

\bibitem[\protect\citeauthoryear{Li}{2017}]{li1}
Li, J., Zheng, W. X., Gu, J., \& Hua, L., 2017, 
Journal of the Franklin Institute, 354(1), 316

\bibitem[\protect\citeauthoryear{Mandel}{202}]{mandel}
Mandel, K., \& Agol, E., 2002, Astrophys. J., 580, 171

\bibitem[\protect\citeauthoryear{Morton}{2016}]{morton}
Morton, T.D., Bryson, S.T., Coughlin, J.L., Rowe, J.F., Ravichandran, G.,
Petihura, E.A., Haas, M.R.,\& Batalha, N. M., 2016,
Ap. J., 822, 86

{ 
\bibitem[\protect\citeauthoryear{Muller}{2013}]{muller}
Muller,  H. M., Huber, K.F., Czesla, S., Wolter, U., \& Schmitt, J. H. M. M., 2013.,
A\&A 560, A112 
}

\bibitem[\protect\citeauthoryear{Murphy}{2012}]{murphy}
Murphy, S.,J., 2012, {\em MNRAS}, 422(1), 665


{  \bibitem[\protect\citeauthoryear{Nutzman}{2009}]{nutz}{
Nutzman, P., Charbonneau, D.,  Winn, J. N., Knutson, H.A., 
Fortney, J.J.,  Holman, M. J., \& Agol, E., 
Ap. J., 692, 229, 2009
}}

\bibitem[\protect\citeauthoryear{Pollacco}{2006}]{Pollacco}
Pollacco, D.L., et al., 2006,  Publ. Astron. Soc. Pac., 118, 1407

\bibitem[\protect\citeauthoryear{Puskullu}{2018}]{puskullu}
Puskullu, C., \& Soydugan, F., 2018, Canadian Journal of Physics, 96, 685.

\bibitem[\protect\citeauthoryear{Raetz}{2014}]{Raetz}
Raetz, St., Maciejeski, G., Ginski, Ch., {\em et al}, 2014, MNRAS, 444, 1351

\bibitem[\protect\citeauthoryear{Rice}{2014}]{rice}
Rice, K., 2014, Challenges, 5, 296

\bibitem[\protect\citeauthoryear{Rhodes}{2014}]{rhodes}
Rhodes, M. and Budding, E., 2014, Ap{\&}SS, 351, 451

\bibitem[\protect\citeauthoryear{Schroter}{2012}]{Schroter}
Schroter, S., Schmitt, J. H. M. M., \& Muller, H. M., 2012, A \& A, 539, 97

{ 
\bibitem[\protect\citeauthoryear{Southworth}{2008}]{Southworth1}
Southworth, J., 2008, MNRAS, 386, 1644
}

{ 
\bibitem[\protect\citeauthoryear{Southworth}{2009}]{Southworth2}
Southworth, J., 2009, MNRAS, 394, 272
}

{ 
\bibitem[\protect\citeauthoryear{Southworth}{2010}]{Southworth3}
Southworth, J., 2010, MNRAS, 408, 1689
}

{ 
\bibitem[\protect\citeauthoryear{Southworth}{2011}]{Southworth4}
Southworth, J., 2011, MNRAS, 417, 2166
} 

{ 
\bibitem[\protect\citeauthoryear{Southworth}{2012}]{Southworth5}
Southworth, J., 2012, MNRAS, 426, 1291
}

{ 
\bibitem[\protect\citeauthoryear{Thompson}{2018}]{Thompson}
Thompson, S. E.,  Coughlin, J. L.,  Hoffman, K.,  Mullally, F., {\em at al.}, 2018,
ApJS, 235, 38
}

\bibitem[\protect\citeauthoryear{Torres}{2008}]{Torres}
Torres, G., Winn, J. N., \& Holman, M. J., 2008, Ap. J, 677, 1324

\bibitem[\protect\citeauthoryear{Turner}{2016}]{Turner}
Turner, J.D., Perason, K.A., Biddle, L. I., {\em et al.},
2016, MNRAS, 459, 789

\end{thebibliography}
 \end{document}